\documentclass{iopart}

\usepackage{graphicx}
\usepackage{iopams}

\begin{document}

\title[Robustness of Topological Order in Semiconductor-Superconductor Nanowires]{Robustness of Topological Order in Semiconductor- Superconductor Nanowires in the Coulomb Blockade Regime}

\author{Bj\"orn Zocher$^{1,2}$, Mats Horsdal$^1$ and Bernd Rosenow$^1$}

\address{$^1$ Institut f\"ur Theoretische Physik, Universit\"at Leipzig, D-04103 Leipzig, Germany\\
$^2$ Max-Planck-Institute for Mathematics in the Sciences, D-04103 Leipzig, Germany}

\begin{abstract}
Semiconductor-superconductor hybrid systems are promising candidates for the  realization Majorana fermions and topological order, i.e. topologically protected degeneracies, in solid state devices. We show that the topological order is mirrored in the excitation spectra and can be observed in nonlinear Coulomb blockade transport through a ring-shaped nanowire. Especially, the excitation spectrum is almost independent of magnetic flux in the topologically trivial phase but acquires a characteristic $h/e$ magnetic flux periodicity in the nontrivial phase. The transition between the trivial and nontrivial phase is reflected in the closing and reopening of an excitation gap. We show that the signatures of topological order are robust against details of the geometry, electrostatic disorder, and the existence of additional subbands and only rely on the topology of the nanowire and the existence of a superconducting gap. Finally, we show that the coherence length in the nontrivial phase is much longer than in the trivial phase. This opens the possibility to coat the nanowire with superconducting nanograins and thereby significantly reduce the current due to cotunneling of Cooper pairs and to enhance the Coulomb charging energy without destroying the superconducting gap. 
\end{abstract}

\date{\today}

% Superconductors transport properties, 74.25.F-
% Single-electron devices, 85.35.Gv
% Mesoscopic systems superconducting, 74.78.Na
% Superconductivity pairing symmetries, 74.20.Rp

%\maketitle

%%%%%%%%%%%%%%%%%%%%%%%%%%%%%%%%%%%%
%%%%%%%%%%%%%%%%%%%%%%%%%%%%%%%%%%%%
\section{Introduction}
\label{sec:introduction}

Topological phases are quantum phases which cannot be described by a local order parameter. Instead, the defining characteristic of topological phases is a pattern of long-range quantum entanglement which is called topological order \cite{W1990,KP06,LW06}. One characteristic property of topological order is the dependence of the ground-state degeneracy on the topology of the manifold on which the system is defined \cite{W1990,RG2000,OKSNT2007}. This degeneracy on manifolds might also serve as a starting point for a general classification of topological phases of strongly correlated quantum matter, complementary to the topological band theory which is based on single-particle states and cannot be easily generalized to correlated systems \cite{WQ2010,G2011}. Recently, there is much interest in topological phases \cite{HK2010,QZ2010} due to their possible application in topological quantum computation. However, these phases are also of fundamental scientific interest for their ability to support exotic quasiparticle (QP) excitations with abelian and even nonabelian quantum statistics. 

One particularly interesting class of topological phases are topological superconductors, which have been predicted to host Majorana bound states \cite{RG2000,FK2009,SF2009,V2009,SLTS2010, LTYSN2010,CF2011,STLSS2011,WSBT2011,B2012,A2012,LF2012b}. The $p_x+ip_y$ superconductor (SC) for spinless fermions is a prototype system for topological SCs. Depending on the chemical potential, the ground state of the $p_x+ip_y$ SC is realized by the weak or the strong pairing phase, which can be distinguished topologically. In particular, the grand canonical ground state of the weak pairing phase on the torus depends on boundary conditions (BCs) for each of the two primitive directions~\cite{RG2000,OKSNT2007}. Here, the ground state with only periodic BCs is special and shows an odd parity, while the three ground states with at least one antiperiodic BC are characterized by an even parity. In contrast, the strong pairing phase and also the ordinary $s$-wave SC on the torus possess a fourfold degenerate even parity ground state~\cite{OKSNT2007}. 

In this paper, we consider a quasi one-dimensional ring shaped SC nanowire and demonstrate that essential aspects of the above described topological degeneracy on the torus carry over to this simpler geometry. We focus on a regime in which the quasiparticle gap $\Delta$ is larger than the single-particle level spacing $d$. In the Coulomb blockade regime, the total particle number and hence the parity of the SC nanowire are fixed by the charging energy $E_c>\Delta$ and the degeneracy of grand canonical ground states is reflected in the excitation energies, which can be observed in nonlinear Coulomb blockade transport~\cite{BRT1996,DR2001}. The lowest excited state above the ground state of a trivial SC with even parity involves two Bogoliubov QPs and thus breaks a Cooper pair, incurring an excitation energy $\delta E \approx 2 \Delta$, which is essentially independent of magnetic flux \cite{ZHR2012}. In contrast, the ground state for odd parity always has one Bogoliubov QP, and hence the lowest excited state involves both annihilating and creating a Bogoliubov QP which costs the excitation energy $\delta E \approx d^2/\Delta \ll 2\Delta$. For nontrivial topological SCs the situation is very different. Here, ground states without unpaired particles at the Fermi energy have odd parity for periodic BC, and even parity for anti-periodic BC. Therefore, the excitation energy $\delta E$ oscillates between $d^2/\Delta$ and $2 \Delta$ as function of magnetic flux with period $h/e$ which is doubled as compared to the case of a trivial SC \cite{ZHR2012}. This connection between the ground-state degeneracy on manifolds with nonzero genus and the $h/e$ flux periodicity of ring structures demonstrates that these properties are a general consequence of topological order and that nonlinear Coulomb blockade transport is a suitable tool to investigate topological order.

Recent experiments show evidence for Majorana fermions and topological superconductivity~\cite{MZ2012,WB2012,RL2012,DY2012,DR2012} in semiconductor (SM) nanowires with strong Rashba spin-orbit coupling in a magnetic field and proximity coupled to an $s$-wave SC~\cite{A2010,LSS2010,ORO2010,AOROF2011}. Despite the effort which has been invested, clear experimental signatures of the unconventional nature of the superconducting state are still missing. Therefore, additional detection schemes have been suggested such as the periodicity of the Josephson effect~\cite{FK2009,K2001,LPAROO2011,HHAB2011}, tunneling spectroscopy~\cite{LLN2009,F2010,WADB2011,LF2011,LB2011}, interferometry~\cite{ANB2009,GS2011}, transport experiments~\cite{TSSZS2011,ADHWB2011,LT2011}, and coincidence measurements~\cite{ZR2010}. We here propose another experiment which directly investigates consequences of topological order on a nontrivial manifold. For this purpose, we use the Coulomb energy as an instrument to prescribe the parity of the hybrid system and thus to observe the above discussed ground-state degeneracy. Our analysis is based on the identification of the pfaffian $\mathbb{Z}_2$ invariant $\mathcal{Q}$ for Hamiltonians in class D \cite{K2001} with the parity of the grand canonical ground state. Thus, we use this key piece of information about the grand canonical ground state to construct two classes of states with parity $\mathcal{Q}$ and $-\mathcal{Q}$, where the class of states with parity $\mathcal{Q}$ ($-\mathcal{Q}$) contains all eigenstates with an even (odd) number of QP excitations. We find two types of excitation spectra which display trivial or nontrivial superconductivity depending on parameter values. The transition between the different topological phases is characterized by the closing and reopening of an excitation gap. As these findings only rely on the existence of a superconducting gap $\Delta>d$ and the $S^1$ topology of the system, the excitation spectra are robust against disorder, spatial variations of the superconducting pairing potential, geometry details, and the existence of additional transverse subbands.

This paper is organized as follows: In section \ref{sec:model_system}, we introduce the model system and the proposed experimental setup. We continue in section \ref{sec:single_band_hamiltonian} with a brief review of the results in Ref. \cite{ZHR2012} for single-band SM nanowires and study the robustness of these results against details of the geometry, electrostatic disorder, and local variations of the superconducting order parameter. In section \ref{sec:multi_band_hamiltonian} we make a departure from the case of strictly one-dimensional nanowires and consider the experimentally realistic case of quasi one-dimensional nanowires. In section \ref{sec:cotunneling_of_cooper_pairs}, we compare the current for the single-electron tunneling with the current due to cotunneling of Cooper pairs which is the most relevant transport channel competing with sequential tunneling of electrons. We summarize our results in section \ref{sec:summary}.

%%%%%%%%%%%%%%%%%%%%%%%%%%%%%%%%%%%%
%%%%%%%%%%%%%%%%%%%%%%%%%%%%%%%%%%%%

\section{Model system}
\label{sec:model_system}

We consider a quasi one-dimensional SM nanowire with strong spin-orbit coupling which is proximity coupled to an $s$-wave SC. The nanowire forms an annulus in the $x$-$y$ plane with radius $R$ and radial extension $L_\perp \ll R$. We assume a strong confinement in $z$-direction, i.e. the extension perpendicular to the plane of the annulus $L_z \ll L_\perp$, such that only the lowest subband with momentum in $z$-direction is occupied. This hybrid system is separated from a back-gate by a thin insulating layer and weakly tunnel-coupled to source and drain electrodes with potentials $\pm eV/2$. Assuming a strong capacitive coupling between the nanowire and the SC, the total number of electrons in this system is determined by the Coulomb Hamiltonian 
%
%***********************  charging energy  *****************
\begin{equation}
H_C = E_c( N + N_{SC} )^2 - eV_G( N + N_{SC} ),
\label{eqn:Charge}
\end{equation}
%************************************************
%
where $E_c$ denotes the charging energy and $N$ ($N_{SC}$) the number of excess electrons in the SM (SC) attracted by the back-gate. Varying the gate potential $eV_G$ allows to change the total electron number $N+N_{SC}$ in discrete units. A current through this island involves changing the electron number from $N+N_{SC}$ to $N+N_{SC} \pm 1$ and creating or annihilating a QP excitation. Thus, resonances in the differential conductivity appear when the condition 
\begin{equation}
eV/2 = E_{N+N_{SC} \pm 1} - E^\mathrm{gs}_{N+N_{SC}}
\end{equation}
is satisfied, where $E_{N+N_{SC}}$ is the total energy of a state with $N+N_{SC}$ electrons and $E^\mathrm{gs}_{N+N_{SC}}$ the respective ground-state energy. The spacing between the resonance peaks is independent of the charging energy $E_c$ and displays the excitation spectrum for fixed particle number, 
\begin{equation}
\delta E_{N+N_{SC}} = E_{N+N_{SC}} - E^\mathrm{gs}_{N+N_{SC}}. 
\label{eqn:energy_diff}
\end{equation}
In our analysis, we assume that both the charging energy $E_c$ and the QP gap $\Delta_{SC}$ in the SC are larger than the effective gap $\Delta_{\mathrm{eff}}$ in the SM. Hence, for small voltages $eV \lesssim E_c,\Delta_{SC}$ all electrons in the SC are paired and unpaired electrons as well as breaking of Cooper pairs can only show up in the SM. As a consequence, the parity in the SC is always even and the parity in the SM nanowire is determined by the total parity of the hybrid system. The Coulomb Hamiltonian Eq. \eref{eqn:Charge} fixes the total particle number while the particle number in the SM nanowire fluctuates because of the strong coupling between the SM nanowire and the SC. In the following, we assume that the average particle number in the SM nanowire is fixed and whenever we refer to the particle number in the SM nanowire, we refer to its average.

We describe the low-energy physics of the nanowire by the lattice Hamiltonian $H=H_{\rm SM}+H_{\rm SC}$~\cite{A2010,LSS2010,ORO2010,AOROF2011} with
%
%***********************  SM Hamiltonian  *****************
\begin{eqnarray}
H_{\rm SM}  =& \sum_{\mathbf{r}, \mathbf{r}', \sigma} \Big\{-t_{\mathbf{r}\mathbf{r}'} + \delta_{\mathbf{r},\mathbf{r}'} \left(E_Z \hat{\sigma}^z_{\sigma\sigma} -\mu +V_\mathbf{r} \right)\Big\}c_{\mathbf{r}\sigma}^\dagger c_{\mathbf{r}'\sigma}  \nonumber \\ 
&+\frac{i \alpha }{2a} \sum_{\mathbf{r},\sigma  }\left( c_{\mathbf{r} \sigma}^\dagger \hat{\sigma}^y_{\sigma \bar{\sigma}} c_{\mathbf{r}+ {\bdelta}_x\bar{\sigma}}-c_{\mathbf{r} \sigma}^\dagger \hat{\sigma}^y_{\sigma \bar{\sigma}} c_{\mathbf{r}- {\bdelta}_x\bar{\sigma}}\right)\nonumber\\
&-\frac{i \alpha }{2a} \sum_{\mathbf{r},\sigma  }\left(c_{\mathbf{r} \sigma}^\dagger \hat{\sigma}^x_{\sigma\bar{\sigma}} c_{\mathbf{r}+ {\bdelta}_y \bar{\sigma}}-c_{\mathbf{r} \sigma}^\dagger \hat{\sigma}^x_{\sigma\bar{\sigma}} c_{\mathbf{r}- {\bdelta}_y \bar{\sigma}} \right), 
\label{eqn:HSM}
\end{eqnarray}
%************************************************
%
where the operator $c_{\mathbf{r}\sigma}^\dagger$ ($c_{\mathbf{r}\sigma}$) creates (annihilates) an electron at site $\mathbf{r}$ with spin $\sigma$ and mass $m$. The first term describes hopping on a simple square lattice with lattice parameter $a$, $t_{\mathbf{r},\mathbf{r}+{\bdelta}} = t_0 \equiv \hbar^2/2m^2a^2$ for the nearest--neighbor lattice vectors ${\bdelta}$, and $t_{\mathbf{r},\mathbf{r}} = -2 t_0$. The second term in Eq. \eref{eqn:HSM} contains the chemical potential $\mu$, the electrostatic disorder potential $V_{\mathbf{r}}$, and the Zeeman energy splitting $E_Z=g\mu_BB/2$ due to the magnetic field in $z$-direction. The last terms in Eq. \eref{eqn:HSM} represent the Rashba spin-orbit coupling with spin-orbit velocity $\alpha$, and $\hat{\sigma}^s$ are the Pauli spin matrices with $s=x,y,z$. By coupling electrons with opposite spins, the spin-orbit coupling creates two helical bands with spin rotating in the $x$-$y$-plane. The magnetic field tilts the spin direction out of the $x$-$y$ plane and removes a level crossing at zero momentum by opening a spin gap.

The proximity coupling between the $s$-wave SC and the nanowire is described by the effective $s$-wave pairing Hamiltonian
%
%***********************  proximity effect  *****************
\begin{equation}
H_{\rm SC} =\sum_{\mathbf{r}}  \Big( \Delta_\mathbf{r} c_{\mathbf{r}\uparrow}^\dagger c_{\mathbf{r}\downarrow}^\dagger   +\Delta_\mathbf{r}^* c_{\mathbf{r}\downarrow}c_{\mathbf{r}\uparrow} \Big)
\label{eqn:HSC}
\end{equation}
%************************************************
%
with pairing potential $\Delta_\mathbf{r}$. The pairing potential modifies the dispersion relation in a crucial way by inducing two excitation energies. It opens an effective pairing gap at the Fermi surface, and reduces the spin gap at zero momentum. Depending on $E_Z$, $\mu$, and $\Delta$, the SM nanowire shows two topologically distinct superconducting phases which are separated from each other by a topological phase transition \cite{A2010,LSS2010,ORO2010}. For one partially occupied subband and $E_Z^2<\Delta^2+\mu^2$, the nanowire is in the topologically trivial phase. By tuning the chemical potential via the gate voltage or the Zeeman energy via the magnetic field, the nanowire can be tuned across the phase transition, which shows up as the closing and reopening of the QP excitation gap. In the topologically nontrivial phase, which is reached for $E_Z^2>\Delta^2+\mu^2$, such a nanowire supports a pair of zero-energy Majorana bound states located at the ends of the wire. 

We here consider a closed nanowire without end points and thus without Majorana bound states. However, the unconventional nature of the topologically nontrivial superconducting phase shows up in the doubling of the magnetic flux period of the excitation spectrum from $h/2e$ to $\Phi_0=h/e$ \cite{ZHR2012}. In our analysis, we incorporate the magnetic flux $\Phi$ through the azimuthal vector potential $\mathbf{A}=\Phi \hat{e}_\varphi/2\pi R$ and the Peierls substitution with 
%
%***********************  Peierls substitution  *****************

\numparts
\begin{eqnarray}
t_{\mathbf{r},\mathbf{r}+{\bdelta}} & \rightarrow t_{\mathbf{r},\mathbf{r}+{\bdelta}} e^{-\frac{ie}{\hbar} \int_{\mathbf{r}}^{\mathbf{r}+{\bdelta}} \mathbf{A}(\mathbf{r}') d\mathbf{r}'}, \\
\alpha & \rightarrow \alpha e^{-\frac{ie}{\hbar} \int_{\mathbf{r}}^{\mathbf{r}+{\bdelta}} \mathbf{A}(\mathbf{r}') d\mathbf{r}'}, \\
\Delta & \rightarrow \Delta e^{i\mathbf{q}\cdot \mathbf{r}},
\label{eqn:Peierls}
\end{eqnarray}
\endnumparts
%************************************************
%
where $\mathbf{q}$ is the Cooper pair wavenumber. We determine the Cooper pair wavenumber by minimizing the ground-state energy of the nanowire with respect to arbitrary vectors $\mathbf{q}$. This is equivalent to determining $\mathbf{q}$ by minimizing the Ginzburg-Landau free energy for the $s$-wave SC coupled to the nanowire \cite{ZHR2012}. For the strictly one-dimensional ring-shaped nanowire with azimuthal $\mathbf{q}=q \hat{e}_\varphi$ this demands that $q$ is the integer nearest to $-2\Phi/\Phi_0$.

We diagonalize the Hamiltonian by defining the Bogoliubov QP operators $\alpha_l=\sum_{\mathbf{r},\sigma} \big(u_{\mathbf{r}\sigma l}c_{\mathbf{r}\sigma}+v_{\mathbf{r}\sigma l}c_{\mathbf{r}\sigma}^\dagger \big)$, where $\{l\}$ is a complete set of QP quantum numbers. This yields 
\begin{equation}
H=\sum_l E_l \alpha_l^\dagger \alpha_l +E_\mathrm{GC}
\end{equation}
with $E_l>0$ and the ground-state energy 
%
%***********************  grand canonical energy  *****************
\begin{equation}
E_\mathrm{GC}= -\frac{1}{2}\sum_l E_l + \frac{1}{2} \sum_{\mathbf{r}, \sigma} \big(2t_0 -\mu + V_\mathbf{r} \big).
\label{eqn:energy_grandcanonical}
\end{equation}
%************************************************
%
The corresponding ground-state electron number is given by the expectation value of the particle number operator $\hat{N}=\sum_{\mathbf{r}\sigma} c_{\mathbf{r}\sigma}^\dagger c_{\mathbf{r}\sigma}$ in the state where all QP levels are empty. Rewriting the particle number operator in terms of QP operators and taking the expectation value with respect to the ground state, we find 
%
%***********************  grand canonical particle number  *****************
\begin{equation}
N_\mathrm{GC}=\sum_{\mathbf{r}\sigma} \sum_l |v_{\mathbf{r}\sigma l}|^2.
\label{eqn:number_grandcanonical}
\end{equation}
%************************************************
%
The parity of the grand canonical ground state is determined by the pfaffian $\mathbb{Z}_2$ topological number 
%
%***********************  topological number  *****************
\begin{equation}
\mathcal{Q}=\frac{\mathrm{Pf}\left( \mathcal{H}i\tau^x \right)}{\sqrt{|\mathrm{det}\left( \mathcal{H}i\tau^x \right)|}} ,
\label{eqn:topol}
\end{equation}
%************************************************
%
where $\mathcal{H}$ denotes the Bogoliubov-de Gennes Hamilton matrix in the basis $(c_{\mathbf{r}\uparrow}^\dagger,\, c_{\mathbf{r}\downarrow}^\dagger,\, c_{\mathbf{r}\uparrow},\, c_{\mathbf{r}\downarrow})$ and $\tau^x$ denotes the Pauli matrix acting on the particle-hole space \cite{K2001,TS2012}. Here, the topological number $\mathcal{Q}=+1(-1)$ corresponds to the even (odd) parity of the grand canonical ground state. From the grand canonical ground state with parity $\mathcal{Q}$ we construct two classes of states with parity $P$ by creating $N_{\mathrm{qp}}$ QP excitations. The parity $P$ of these states is determined by
\begin{equation}
P=\mathcal{Q} \, \cdot (-1)^{N_{\mathrm{qp}}},
\end{equation}
i.e. depending on whether $N_{\mathrm{qp}}$ is even (odd), $P=\mathcal{Q}$ ($P=-\mathcal{Q})$).

Since the large Coulomb energy enforces a well-defined parity and mean particle number in the nanowire, we Legendre transform both classes of states into a pseudo-canonical ensemble with $N_{\mathrm{qp}}$ QPs, mean particle number $N$, and parity $P$. This ensemble contains states with energy
%
%***********************  canonical energy  *****************
\begin{equation}
E[\{l_i\},N,P] = E_\mathrm{GC} + \sum_{j=1}^{N_{\mathrm{qp}}} E_{l_j} +\mu N,
\label{eqn:energy_canonical}
\end{equation}
%************************************************
%
where the chemical potential $\mu$ is determined by the constraint 
%
%***********************  canonical particle number  *****************
\begin{equation}
N = N_\mathrm{GC}(\mu)+\sum_{j=1}^{N_{\mathrm{qp}}} \sum_{\mathbf{r}\sigma} \big(|u_{\mathbf{r}\sigma l_j}(\mu)|^2-|v_{\mathbf{r}\sigma l_j}(\mu)|^2 \big).
\label{eqn:number_canonical}
\end{equation}
%************************************************
%
Here, $\{ l_i \, : i=1,\,  \dots, N_\mathrm{qp} \}$ denotes the set of quasiparticle excitations. For noninteracting systems, this method is fully equivalent to the wave function based technique used in Ref. \cite{ZHR2012}. There, the authors defined two classes of grand canonical ansatz wave functions with even and odd parity as tensor products of generalized BCS wave functions \cite{DR2001}. The ground state was then determined by the unbiased minimization of the energy expectation value and the lowest excited states where given by pairwise creation of Bogoliubov QPs. However, while the wave function method is defined on the many-body Hilbert space, the technique used here only relies on the pfaffian $\mathbb{Z}_2$ invariant, the QP energies, and the corresponding QP wave functions and is thus more suitable for large systems without additional symmetries. 

In recent experiments, it has been shown that both InSb and InAs are suitable SM materials due to a large $g$-factor and strong spin-orbit coupling~\cite{MZ2012,WB2012,RL2012,DY2012,DR2012}. In the experimental situation, the confinement energy in transverse direction is the largest energy scale so that only a few subbands are partially occupied. It is useful to express the kinetic energy and the spin-orbit coupling in terms of a characteristic energy $\epsilon_{so}=m\alpha^2 /\hbar^2$ and a spin-orbit length $l_{so}=\hbar^2/m\alpha$. Typical values for these materials are $\epsilon_{so}=0.1$ meV, $l_{so}=100$ nm, and $g\mu_B/2=1\, \mathrm{meV/T}$. Thus, we find with $R=0.5 \, \mu \mathrm{m}$, $\Delta=0.5$ meV, and $E_Z=1$ meV, an effective pairing gap of $\Delta_{\mathrm{eff}} \approx 0.2\, \mathrm{meV}$ and the level spacing at the Fermi energy of $d = 0.08 \, \mathrm{meV}$. To ensure sequential single-electron tunneling through the hybrid system, we consider the case $E_c \gg\Delta_{\mathrm{eff}}$. Here, a relative charging energy between the SM and the SC of the order of 1 meV together with a pairing gap $\Delta_{SC}=2$ meV in the SC~\cite{MZ2012} would reduce $\Delta_{\rm eff} $ by 20 \% \cite{OS2004} and is thus neglected. In the following, we vary the magnetic field $B$ in discrete steps with the magnetic flux always being fixed modulo $\Phi_0$, such that the only effect of $B$ is a change of the Zeeman energy $E_Z$.

%%%%%%%%%%%%%%%%%%%%%%%%%%%%%%%%%%%%
%%%%%%%%%%%%%%%%%%%%%%%%%%%%%%%%%%%%

\section{Single-band Hamiltonian}
\label{sec:single_band_hamiltonian}

We begin our analysis with a perfectly one-dimensional nanowire of width $L_\perp\rightarrow 0$ which has been studied recently in Ref. \cite{ZHR2012}. There, the authors considered the rotational symmetric continuum version of the model introduced above, i.e. spatially constant superconducting pairing $\Delta_{\mathbf{r}}=\Delta$ and vanishing electrostatic disorder $V_{\mathbf{r}}=0$. Before discussing the relevance of geometry details, electrostatic disorder, and spatial dependence of the pairing potential, we recapitulate the main results of Ref. \cite{ZHR2012}. To illustrate the main results in that Letter, we will here focus on the non-disordered case (black line  labeled by "$\gamma=0$") in figures \ref{fig:flux_singleband} and  \ref{fig:field_singleband}. The disordered case will be treated in section \ref{sec:electrostatic_disorder}. The plot shows the lowest excitation energy $E_N-E_N^{gs}$ as a function of Zeeman energy with $N$ chosen such that $\mu \approx 0$. As discussed above, at $E_Z^2=\Delta^2+\mu^2$ the Zeeman energy drives the nanowire through a topological phase transition with the trivial phase for $E_Z\lesssim \Delta$ and the nontrivial phase for $E_Z \gtrsim \Delta$~\cite{STLSS2011}.

Our findings for $E_Z\lesssim \Delta$ are characteristic for $s$-wave superconductivity in metallic nanograins~\cite{BRT1996, DR2001}. For even parity, the excitation spectrum shows a superconducting gap $2 \Delta_{\mathrm{eff}}$ since all excited states contain two Bogoliubov QPs which corresponds two breaking one Cooper pair. In contrast, the ground state for odd parity always has one Bogoliubov QP and therefore the spectrum is qualitatively independent of magnetic flux and determined by the single-particle level spacing as 
\begin{equation}
\delta E = E(n=1) - E(n=0) \approx \frac{d^2}{2 \Delta_{\mathrm{eff}}}
\end{equation}
for QP energies $E(n) = \sqrt{n^2 d^2 + \Delta_{\mathrm{eff}}^2}$ where $n$ counts the energy levels relative to the Fermi energy. Similarly, a variation of magnetic flux by $\Phi_0/2$ changes $E(n)$ on the order of $d^2/\Delta_{\mathrm{eff}}$. 

In the topologically nontrivial phase for $E_Z\gtrsim \Delta$, the parity effect is very different. Here, the excitation energies depend on both electron parity and magnetic flux. In Figs.~\ref{fig:field_singleband}(a) and (d) we find a QP excitation gap $2\Delta_{\mathrm{eff}}$ since two Bogoliubov QP excitations are required and thus a Cooper pair needs to be broken. In contrast, the excitation energies in Figs.~\ref{fig:field_singleband}(b) and~\ref{fig:field_singleband}(c) reflect the single-particle level spacing as $d^2/2 \Delta_{\mathrm{eff}}$ since always one unpaired particle is located near the Fermi surface. As shown in Fig.~\ref{fig:field_singleband}(d), the characteristic signature of the topological phase transition at $E_Z \sim \Delta$ is the closing and reopening of the QP excitation gap. 

These different parity effects become even more impressive when fixing the Zeeman energy and varying the magnetic flux. In the trivial phase, the excitation energies for even parity are of order $2 \Delta_{\mathrm{eff}}$ with small oscillations of period $\Phi_0/2$ and amplitude $d^2/\Delta_{\mathrm{eff}}$. For odd parity, the QP gap is absent and the excitation energies reflect the single-particle level spacing. In contrast, as shown in figure~\ref{fig:geometry}(b) we find large $\Phi_0$ periodic oscillations of amplitude $2 \Delta_{\mathrm{eff}}$ in the nontrivial phase. Here, the excitation energies for $\Phi/\Phi_0 \in (-1/4,1/4)$ are determined by the superconducting gap $2\Delta_{\mathrm{eff}}$ due to the pairwise creation of Bogoliubov QPs i.e. the breaking of Cooper pairs, while they display the single-particle level spacing $d^2/\Delta_{\mathrm{eff}}$ for $\Phi/\Phi_0 \in (1/4,3/4)$. The excitation spectrum for odd parity is equivalent to that for even parity but shifted by $\Phi_0/2$, as follows from the discussion above.

This characteristic $\Phi_0$ flux periodicity of the excitation spectrum in the nontrivial superconducting phase is directly related to the $4 \pi$ periodicity of the Josephson current between two topological SCs~\cite{FK2009,K2001,HHAB2011} which has been recently discovered in InSb/Nb nanowire junctions \cite{RL2012}. In first order in the tunneling matrix element $t$, the Josephson Hamiltonian between two one-dimensional topological SCs is given by
%
%************************** topological Josephson Hamiltonian ***********************
\begin{equation}
H_J(\Delta\phi)= \mathcal{P} \, t \cos\Big(\frac{\Delta \phi}{2}\Big),
\label{eqn:HJ}
\end{equation}
%******************************************************************************************
%
where $\mathcal{P}$, which has eigenvalues $\pm 1$, describes the parity operator of the neutral fermion state shared between the two topological SCs and $\Delta \phi$ the superconducting phase difference. For a fixed parity of the neutral fermion, the Josephson energy is $4 \pi$ periodic in the phase difference. If the Josephson junction is inserted into a ring structure, the magnetic flux threading the ring yields a superconducting phase difference $\Delta \phi = 4\pi \Phi/\Phi_0$, and the $4 \pi $ phase periodicity is equivalent to a $\Phi_0$ flux periodicity. If the nanowire is coupled to a reservoir, varying $\Delta \phi$ by $2\pi \sim \Phi_0/2$ will change the occupancy $(\mathcal{P} +1)/2$ of the neutral fermion state and hence the parity of the ground state. This is in agreement with our result that the parity of the ground-state wave function changes when varying the flux through the ring by $\Phi_0/2$.

%%%%%%%%%%%%%%%%%%%%%%%%%%%%%%%%%%%%

\subsection{Dependence on details of the geometry}
\label{sec:dependence_on_geometric_details}

\begin{figure}[tb]
\begin{center}
\includegraphics[width=0.62\textwidth ,clip]{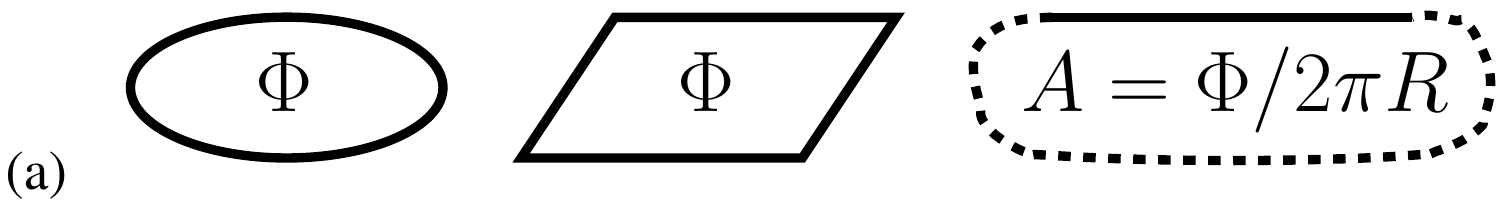}\\
\includegraphics[width=0.62\textwidth ,clip]{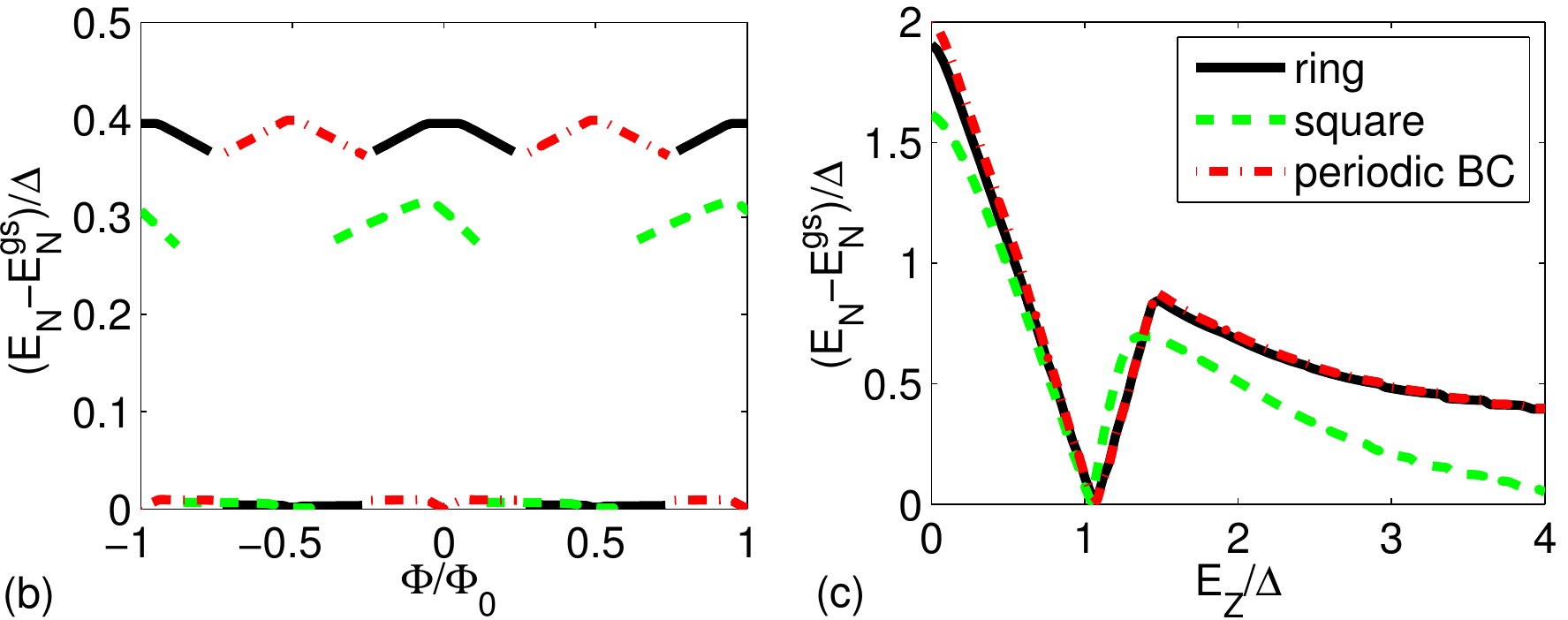}
\end{center}
\caption{(a) Sketch of three different realizations of the $S^1$ topology of the circle: a ring, a square, and a wire with periodic boundary condition indicated by dots. (b) Lowest excitation energies for $E_Z=2\Delta$ as function of magnetic flux, and (c) lowest excitation energies for $\Phi=0$ (mod $\Phi_0$) as function of magnetic field. The parameters used in the calculation are: even parity, $N=44$, $L=2\pi R=3$ $\mu$m, and $a=2$ nm. In (c), the magnetic flux for the system with periodic BC is $\Phi=-\Phi_0/2\, (\mathrm{mod}\, 2)$. The lines for panel (b) are defined in panel (c). } 
\label{fig:geometry}
\end{figure}

In this section, we study how details of the geometric realization of the ring topology affect the excitation spectrum. For this purpose, we compare the spectra for a ring, a square, and for a model with periodic BC as sketched in figure \ref{fig:geometry}(a). In Ref. \cite{ZHR2012}, the authors showed that the low-energy physics of the ring-shaped nanowire is equivalent to that of a strip of width $L_\perp$ and length $L=2\pi R \gg L_\perp$ with periodic BC along the $x$-direction and with vector potential $\mathbf{A}=(\Phi-\Phi_0/2)\hat{x}/L$ in the Landau gauge. Here, the first term in the bracket describes the magnetic flux penetrating the ring. The second term originates from the conservation of the total angular momentum of the electrons in the ring-structure. More specifically, this conservation yields a spin-orbit coupling between states with equal total angular momentum $(pR, \uparrow)$ and $(pR+\hbar,\downarrow)$ and thus effectively shifts orbital angular momenta $pR$ by $\pm \hbar/2$~\cite{MMK2002}. This shift can be understood by identifying it with a $2\pi$ spin rotation of an electron encircling the ring which is equivalent to a Berry phase factor $-1$ and an effective shift of the magnetic flux by $-\Phi_0/2$. 

Experimentally, the fabrication of ring structures with radii of several hundred nanometers is challenging and the approximation of the ring by a triangle or a rectangle is likely. For this purpose, we consider a square-like structure made out of four nanowires, however, our results are qualitatively also valid for a triangular structure which consists of three wires. 

In figure \ref{fig:geometry}(b),(c), we compare the fixed particle number excitation spectrum for the three different realizations of the lattice on which the Hamiltonian is defined; the ring of radius $R$, a quadratic approximation of the ring with edge lengths $\pi R/2$, and a straight nanowire of length $L=2\pi R$ with periodic BC. In our numerics, we model the ring-shaped nanowire by a one-dimensional tight-binding Hamiltonian Eq. \eref{eqn:HSM} with spin-orbit coupling perpendicular to the nanowire, and thus rotating in the $x$-$y$ plane. As function of the discretized azimuthal angle $\varphi_i=2\pi i/n$ with $n$ lattice sites, the spin-orbit direction is then given by
\begin{equation}
\sigma_{\mathrm{SO, ring}}(i) = \sin(\varphi_i) \sigma^x+\cos(\varphi_i)\sigma^y. 
\end{equation} 
Similarly, we model the square by abrupt changes in the spin-orbit direction at the position of the corners, 
\begin{equation}
\sigma_{\mathrm{SO, square}}(i) = \left\{\begin{array}{cl} 
\sigma^x, & \mbox{for }0 < i \le \frac{n}{4}\\
\sigma^y, & \mbox{for }\frac{n}{4} < i \le \frac{n}{2}\\
-\sigma^x, & \mbox{for }\frac{n}{2} < i \le \frac{3n}{4}\\
-\sigma^y, & \mbox{for }\frac{3n}{4} < i \le n
\end{array}\right. 
\end{equation}
and the straight nanowire with periodic BC by a constant spin-orbit direction, 
\begin{equation}
\sigma_{\mathrm{SO, periodic}}(i) = \sigma^y. 
\end{equation} 

The spectra for these three models are qualitatively very similar and show a $\Phi_0$ flux period. As expected, we find that the model with periodic BC yields the same spectrum as the ring model but with vector potential shifted by $-\Phi_0\hat{x}/2L$ due to the Berry phase factor $-1$ which is exact up to corrections of the order of $d/\sqrt{E_Z^2+\alpha^2 k^2} \ll 1$. This phase shift originates from the $2\pi$ spin rotation, and therefore  also exists for the square model where the spin rotation happens in discrete jumps rather than continuously. Without superconductivity $\Delta=0$, the spectra for the ring and the square are identical since both Hamiltonians can be transformed into each other by a local gauge transformation with different gauge fields for spin up and spin down electrons. However, for $\Delta \neq 0$ this transformation is not possible since the spin singlet pairing Hamiltonian breaks the local gauge symmetry. As a consequence, the spectrum for the square geometry shows small deviations from that for the ring geometry because of the existence of corners where the spin-orbit direction jumps by $\pi/2$. In particular, we find that the excitation spectrum for the square is slightly shifted towards smaller values of the magnetic flux as compared to the ring geometry, with the shift being of the order of $\Delta_{\mathrm{eff}}/\alpha k$. In addition to the non-universal phase shift, we find that QP states with reduced excitation energy exist, which are predominately localized near the corners of the square.

We see that our main results are robust against the details of the geometric realization and rely on the existence of a void such that the topology of the nanowire is homotopically equivalent to an annulus. All these results underline the general arguments in the introduction, connecting ground-state degeneracies on the torus to parity and flux periodicities of excitations.

%%%%%%%%%%%%%%%%%%%%%%%%%%%%%%%%%%%%

\subsection{Electrostatic disorder}
\label{sec:electrostatic_disorder}

\begin{figure}[t]
\centerline{\includegraphics[width=0.68\textwidth ,clip]{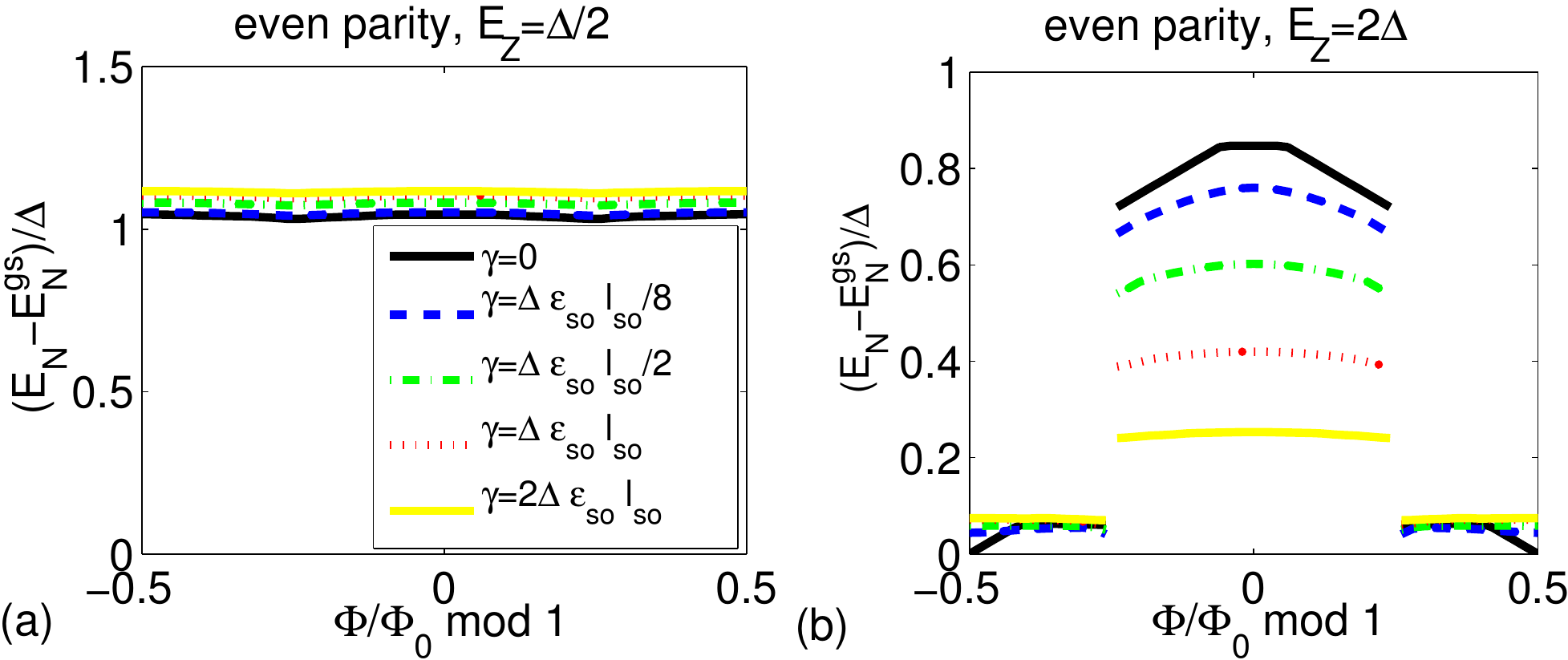}}
\caption{Lowest excitation energy for a ring-shaped nanowire with $N=44$, $L=3$ $\mu$m, and $a=5$ nm as function of magnetic flux and for different variances $\gamma$ of the electrostatic disorder potential. The curves represent the average over 50 random disorder configurations. The lines are defined in panel (a).} 
\label{fig:flux_singleband}
\end{figure}

On one hand, disorder is known to often have drastic influence on the electronic properties of low-dimensional systems. On the other hand, superconducting pairing correlations in $s$-wave SCs are protected against time-reversal invariant impurity scattering by Anderson's theorem \cite{A1959}. This motivates us to address the question of how robust the ground-state degeneracies in the torus topology are against potential electrostatic disorder. In the following, we discuss the effect of disorder on the excitation energies in the regime where the effective gap is larger than the single-particle level spacing, i.e. for 
\begin{equation}
\Delta_{\mathrm{eff}} \equiv \frac{\Delta \epsilon_{so}l_{so}N}{2R E_Z} > d \ \ . 
\end{equation}
We model electrostatic disorder by a locally varying impurity potential $V_{\mathbf{r}}$ with vanishing mean value and Gaussian white noise correlator $\langle V_{\mathbf{r}} V_{\mathbf{r}'} \rangle = \gamma \delta_{\mathbf{r},\mathbf{r}'}/a$. We here consider the regime of disorder strengths $\gamma \lesssim  \gamma_m$ with $\gamma_m = \Delta \epsilon_{so} l_{so}$, since strong disorder $\gamma \gg  \gamma_m$ breaks the nanowire into topological and nontopological domain walls and thereby destroys the excitation gap \cite{F2010}.

\begin{figure}[t]
\centerline{\includegraphics[width=0.68\textwidth ,clip]{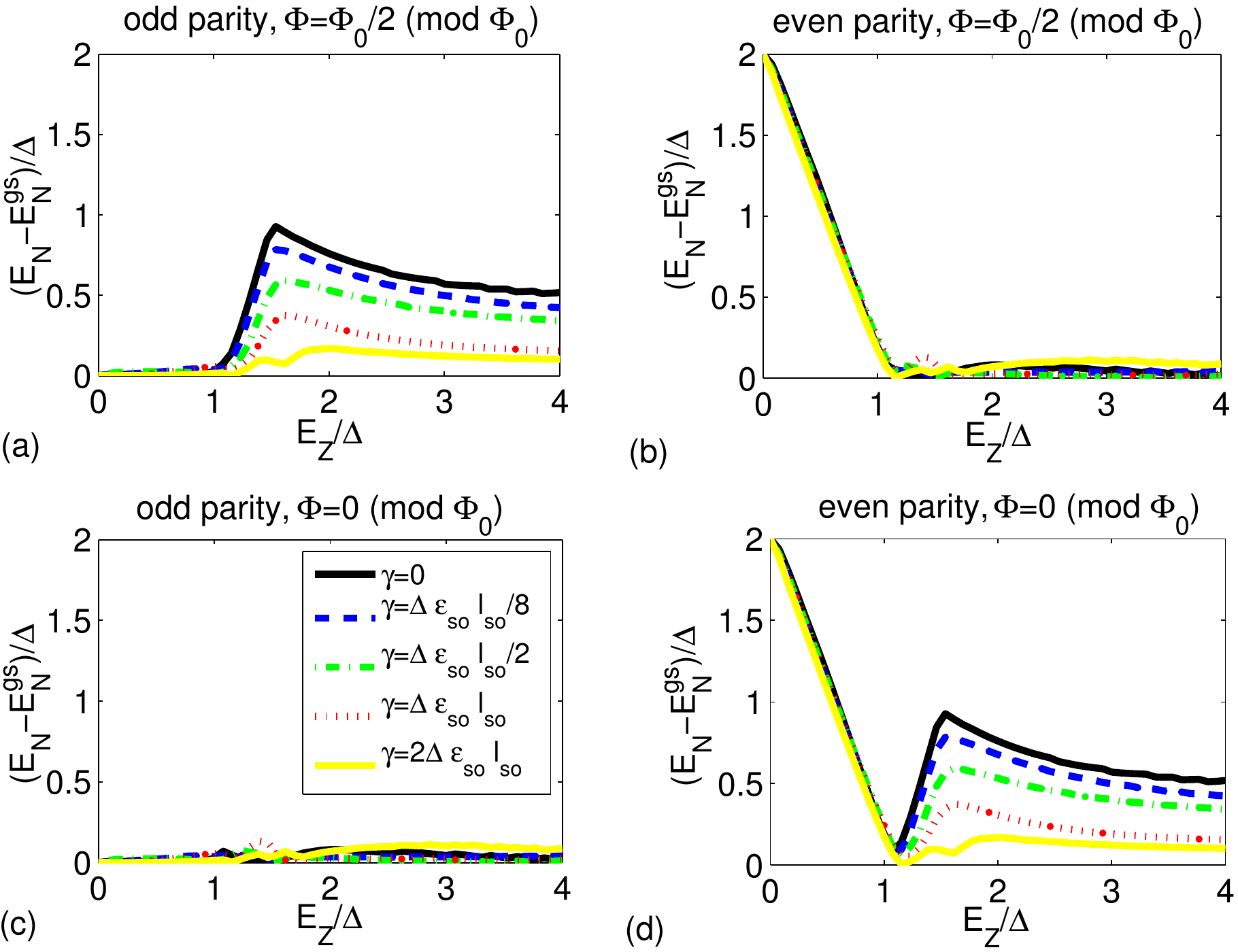}}
\caption{Lowest excitation energy for a ring-shaped nanowire with $N=44$, $L=3$ $\mu$m, and $a=5$ nm as function of Zeeman energy and for different variances $\gamma$ of the electrostatic disorder potential. The curves represent the average over 50 random disorder configurations. The lines are defined in panel (c). } 
\label{fig:field_singleband}
\end{figure}

In figure \ref{fig:flux_singleband} we compare the excitation spectra for the topologically non-trivial and trivial phase as a function of magnetic flux and in figure \ref{fig:field_singleband} we display the excitation spectrum as a function of Zeeman energy for different combinations of parity and magnetic flux. We find that the effect of electrostatic disorder is very different in the topologically trivial and the nontrivial phase. In the trivial phase, the QP excitation gap is remarkably robust against disorder which is characteristic for $s$-wave superconductivity \cite{A1959}. In contrast, we find a significant reduction of the excitation gap due to disorder in the nontrivial phase. While the $\Phi_0$ flux periodicity is not directly affected by disorder as shown in figure \ref{fig:flux_singleband}(b), there is a maximum level of electrostatic disorder $\gamma_m = \Delta \epsilon_{so} l_{so}$ such that only for $\gamma \lesssim \gamma_m$ the excitation gap $\approx 2 \Delta_{\rm eff}$ is larger than the single-particle level spacing $d$ and the $\Phi_0$ periodicity is observable. Since the topological phase for large Zeeman energies $E_Z \gg \Delta,\mu$ can be mapped onto a spinless $p$-wave SC \cite{LSS2010}, this reduction is in full agreement with the effect of disorder on the excitation gap in spinless $p$-wave SCs \cite{MD2001,Brouwer+11}. Furthermore, we find that the reduction is very efficient near the topological phase transition since there already weak disorder breaks the nanowire into domains of different chemical potential and thereby shifts parts of the wire through the topological phase transition which reduces the excitation gap locally. Away from the topological phase transition, the reduction of the excitation gap is weaker because the existence of partially trivial domains due to disorder becomes unlikely. Furthermore, we find that disorder shifts the topological phase transition towards larger values of the Zeeman energy \cite{Brouwer+11,LSS2011}. As before, we argue that this shift originates from local topological phase transitions at $E_Z^2=\Delta^2+(\mu +V_{\mathbf{r}})^2$ which are shifted towards larger Zeeman energies due to disorder.

Since the parity and flux dependence of excitation energies reflect the presence or absence of nontrivial topological order, our findings for the nonlinear Coulomb blockade transport are robust against electrostatic disorder and other perturbations as long as the topological order is not destroyed by the formation of domain walls. In particular, we find a maximum variance  $\gamma_m $ of electrostatic disorder below which the condition $\Delta_{\mathrm{eff}}>d$ is clearly satisfied, and the $\Phi_0$ periodicity is observable.

%%%%%%%%%%%%%%%%%%%%%%%%%%%%%%%%%%%%

\subsection{Nonsuperconducting segments}
\label{sec:non_superconducting_segments}

\begin{figure}[t]
\begin{center}
\includegraphics[width=0.58\textwidth ,clip]{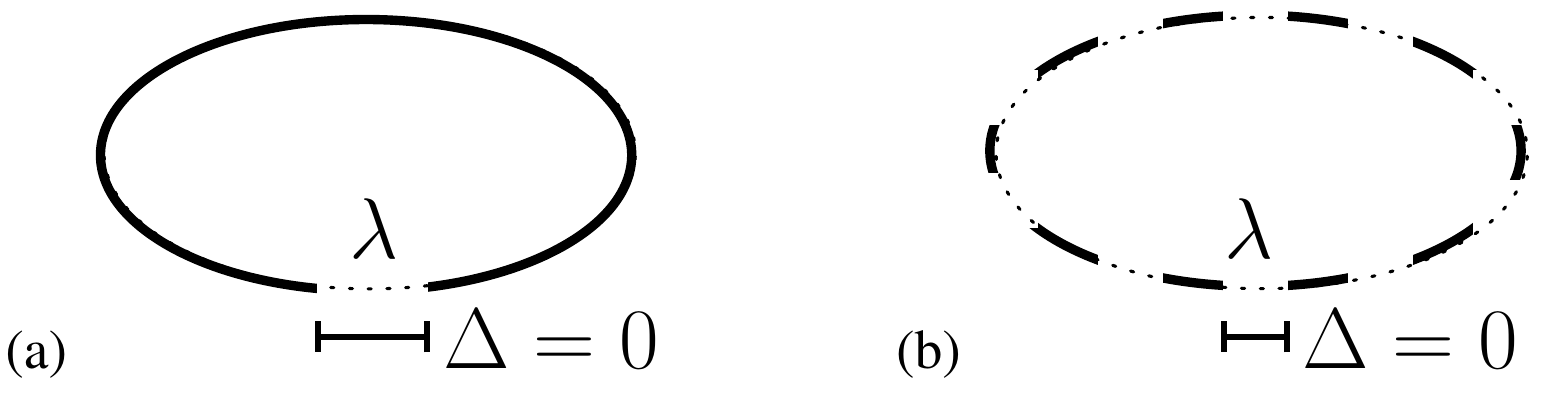}\\
\includegraphics[width=0.62\textwidth ,clip]{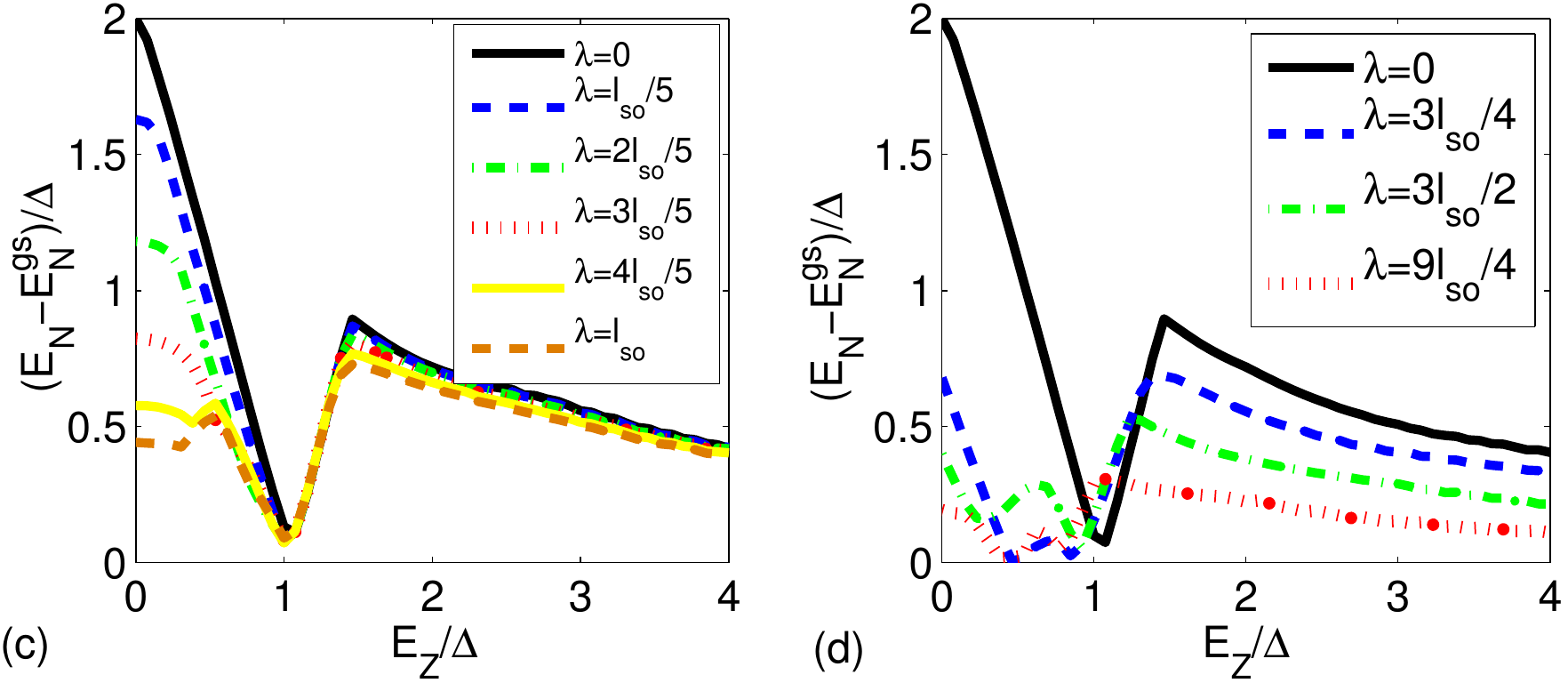}
\end{center}
\caption{Sketch of the nanowire with (a) one and (b) ten nonsuperconducting segments of length $\lambda$, i.e. segments which are not covered with superconducting material. (c), (d) Lowest excitation energy for a ring-shaped nanowire of length $L=3$ $\mu$m with even parity, $N=44$, and $\Phi=0$ for the case with (c) one and (d) ten nonsuperconducting segments. } 
\label{fig:spectrum_open}
\end{figure}

In this section, we consider the situation that the proximity induced superconducting order parameter is spatially dependent. Experimentally this might appear due to the roughness of the nanowire/SC interface or if the nanowire is not completely covered with the $s$-wave SC. As sketched in figure \ref{fig:spectrum_open}(a), we describe this spatial dependence of the superconducting pairing amplitude by a step function such that $\Delta_{\mathbf{r}}=0$ for $0<x<\lambda\ll L$ and $\Delta_{\mathbf{r}}=\Delta$ elsewhere. In figure \ref{fig:spectrum_open}(c), we display the excitation energies for different lengths $\lambda$ of the nonsuperconducting segment. We find a significant reduction of the excitation gap in the trivial phase while the excitation energies in the nontrivial phase are only weakly reduced even for $\lambda \sim l_{so}$. We argue that the origin of the robustness of the excitation gap in the topological phase is the small effective gap $\Delta_{\mathrm{eff}}<\Delta$ and the enhanced Fermi velocity $v_F(E_Z>\Delta)\sim 2 v_F(E_Z=0)$ due to the occupation of a single spinless band. Hence, the superconducting coherence length $\xi =v_F/\Delta_{\mathrm{eff}}$ in the topologically nontrivial phase is significantly enhanced as compared to the trivial phase. Thus, in the nontrivial phase superconducting pairing correlations are more efficiently induced in the nonsuperconducting part of the wire which here shows up as the robustness of the superconducting gap against the existence of a quite long normal segment.

From the robustness of the excitation gap, we conclude that it is not necessary for our proposed setup that the nanowire is completely covered with the $s$-wave SC. In particular, we propose that it is sufficient to place superconducting grains on the nanowire in order to significantly increase the charging energy and to reduce Cooper pair cotunneling through the SC [see section \ref{sec:cotunneling_of_cooper_pairs}]. We now assume that the nanowire contains ten nonsuperconducting segments of length $\lambda$ uniformly distributed over the nanowire as sketched in figure \ref{fig:spectrum_open}(b). In figure \ref{fig:spectrum_open}(d), we study the excitation spectrum for different characteristic lengths $\lambda$. While the excitation gap for $\lambda =3l_{so}/2\sim 150$ nm in the trivial phase is completely absent, we find that the excitation energies in the nontrivial phase are only reduced by 30\% as compared to the situation where $\Delta \neq 0$ everywhere.

There is a renormalization of $\Delta_{\mathrm{eff}}$ in the case of covering the SM nanowire with superconducting grains. Such a  situation was discussed by van Heck {\it et al.} in Ref. \cite{HHAB2011}, and the mechanism for the renormalization of $\Delta_{\mathrm{eff}}$ are phase fluctuations in the regions between two grains, which are enhanced  by the  existence of a relative charging energy between the superconducting grains. The dimensionless parameter controlling the strength of phase fluctuations is $\delta/\Delta_{\mathrm{eff}}$, where $\delta$ denotes the  energy for charging one grain relative to the other. For a covering with distance between the grains much smaller than the coherence length $\xi$, it is reasonable to assume that $\delta \ll E_c$ such that a regime with $\delta < \Delta_{\mathrm{eff}}$ can be reached, where the renormalization of $\Delta_{\mathrm{eff}}$ is unimportant.

%%%%%%%%%%%%%%%%%%%%%%%%%%%%%%%%%%%%
%%%%%%%%%%%%%%%%%%%%%%%%%%%%%%%%%%%%

\section{Multi-band Hamiltonian}
\label{sec:multi_band_hamiltonian}

In this section we make a departure from the case of strictly one-dimensional nanowires and consider the experimentally realistic situation of quasi one-dimensional nanowires of finite thickness with $a\ll L_\perp<\xi$. In our numerical analysis, we model the ring shaped nanowire by a strip with periodic boundary conditions along the $x$-direction and with hard wall boundary conditions along the $y$-direction. The magnetic flux $\Phi$ is incorporated through the modified vector potential $\mathbf{A}=(\Phi-\Phi_0/2)\hat{x}/L$ as discussed above in section \ref{sec:dependence_on_geometric_details}. 

\begin{figure}[t]
\begin{center}
\includegraphics[width=.7\textwidth ,clip]{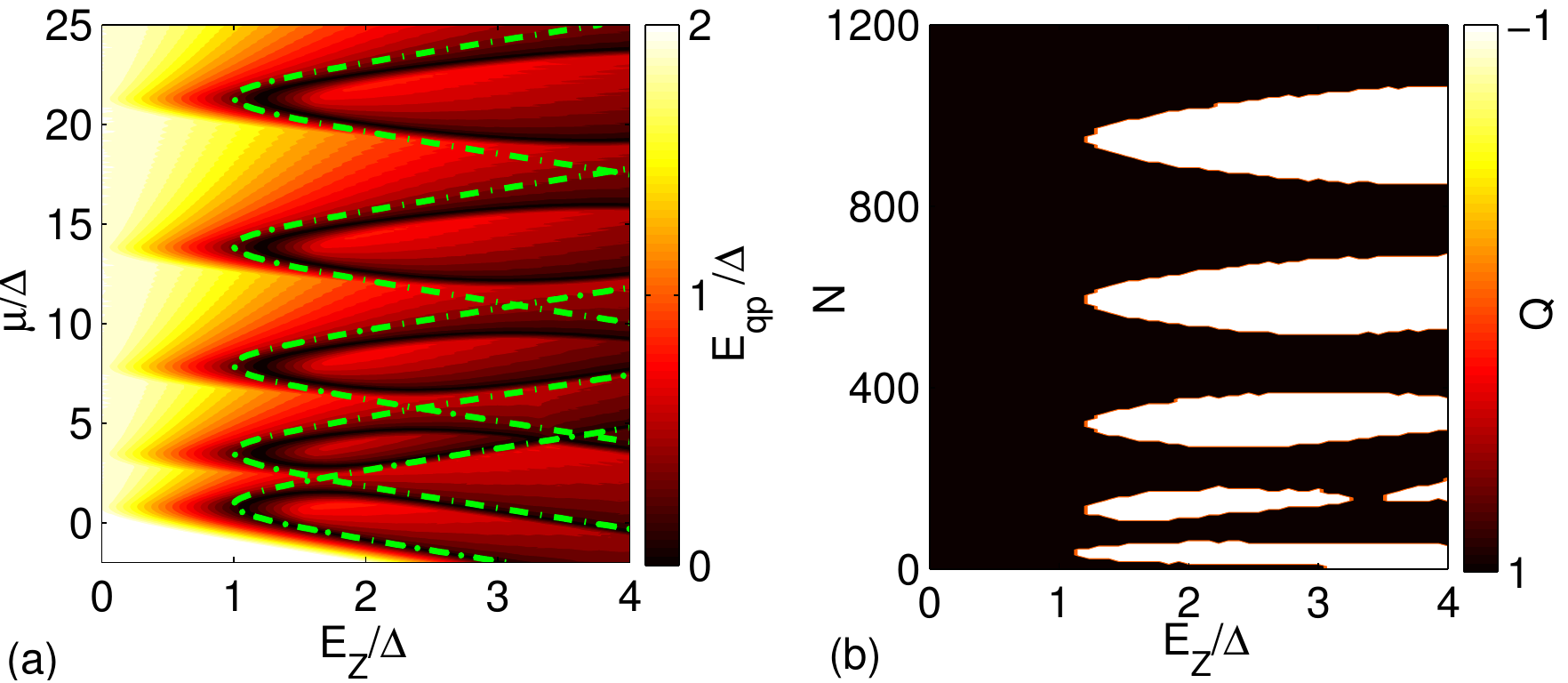}
\end{center}
\caption{(a) Lowest energy of the Bogoliubov QP spectrum $E_{\mathrm{qp}}=\min \{E_l\}$ as function of Zeeman energy and chemical potential, and (b) phase diagram inferred from the topological number $Q$ as function of Zeeman energy and mean electron number. The green dashed lines in (a) represent the topological phase transition for the various subbands in the limit of uncoupled subbands $\alpha \rightarrow 0$, whereas the color scale encodes the excitation energy for coupled subbands. The parameters used in the calculation are: $L=3$ $\mu$m, $L_\perp=100$ nm, $a=5$ nm, and $\Phi=-\Phi_0/2$. } 
\label{fig:spectrum_multiband}
\end{figure}

To ensure that the induced superconducting phase remains quasi one-dimensional and the nanowire exhibits a substantial gap, we demand that the width does not exceed the superconducting coherence length $\xi=v_F/\Delta_{\mathrm{eff}}$ \cite{LSS2011,WA2010,PL2010,ZS2011,LL2011,LF2011a}. The spatial extension in the $y$-direction gives rise to the existence of additional transverse modes and thus subbands which might be partially occupied depending on the chemical potential. In figure \ref{fig:spectrum_multiband}(a), we display the Bogoliubov QP spectrum for $\Phi=-\Phi_0/2$ as function of Zeeman energy and chemical potential. For $\mu \lesssim (\pi\hbar)^2/2mL_\perp^2$ only one subband is partially occupied and the excitation spectrum is equivalent to the one discussed in section \ref{sec:single_band_hamiltonian}. With increasing chemical potential higher subbands are filled up consecutively and similarly to the single-band case, the higher subbands can be either topologically trivial or nontrivial depending on the chemical potential and the Zeeman energy.

In figure \ref{fig:spectrum_multiband}(a), the topologically nontrivial phase shows up as islands which are enclosed by lines of vanishing excitation energies, i.e. by topological phase transitions. Assuming that the subbands are uncoupled, we find the topological phase whenever the chemical potential lies well within one of the spin gaps at zero momentum and when the Zeeman energy satisfies the relation 
\begin{equation}
E_Z^2>\Delta^2+(\mu-\epsilon_n)^2,
\end{equation}
where $\epsilon_n=(\hbar n\pi)^2/2mL_\perp^2$ denotes the kinetic energy of subband $n$. However, the transverse spin-orbit term $\alpha c_{\mathbf{r},\sigma}^\dagger \sigma^x_{\sigma\sigma'}c_{\mathbf{r}+\bdelta_y,\sigma'}$ couples the subbands and thereby modifies the lines where the topological phase transitions occur. These modifications are similar to avoided crossings with energy splitting $\delta \mu \approx 2\alpha k_{F,y}$ between the lines of topological phase transitions and thus reduce the size of the topologically nontrivial islands. With increasing chemical potential, the spin-orbit energy in the transverse direction increases and thus the energy splitting due to the avoided crossing increases $\delta \mu \sim  \alpha \sqrt{\epsilon_n}$.

In figure \ref{fig:spectrum_multiband}(b), we display the topological number $Q$ as function of Zeeman splitting and mean electron number $N$. As before, the topological number is $+1$ in the trivial and $-1$ in the nontrivial phase and thus we conclude that the parity of the grand canonical ground states in both phases is different with even parity in the trivial and odd parity in the nontrivial phase. Similarly to figure \ref{fig:spectrum_multiband}(a), we find islands of topologically nontrivial phase which are enclosed by the trivial phase. We propose that the fixed mean particle number excitation energies can be used as a tool to investigate the topological phase diagram. In \ref{sec:multiband_phase_diagram} we show that our results for the single-band nanowire can directly be applied to the multi-subband nanowire. Thus, the topologically trivial phase is characterized by an excitation spectrum with gap $2\Delta_\mathrm{eff}$ for even parity while the excitations are determined by the single-particle level spacing for odd parity. When varying the magnetic flux, the spectra for both even and odd parity show small $\Phi_0/2$ periodic oscillations as expected for trivial SCs. In contrast, the situation is different in the topologically nontrivial phase where the excitation spectrum qualitatively depends on both magnetic flux and electron parity. We here find a characteristic $\Phi_0$ flux period similar to the situation for the single-band model in section \ref{sec:single_band_hamiltonian}. We conclude that the excitation spectrum for fixed mean particle number, which can be observed in nonlinear Coulomb blockade transport, is an unbiased tool to map out the topological phase diagram shown in figure \ref{fig:spectrum_multiband}(b).

\begin{figure}[t]
\begin{center}
\includegraphics[width=0.4\textwidth ,clip]{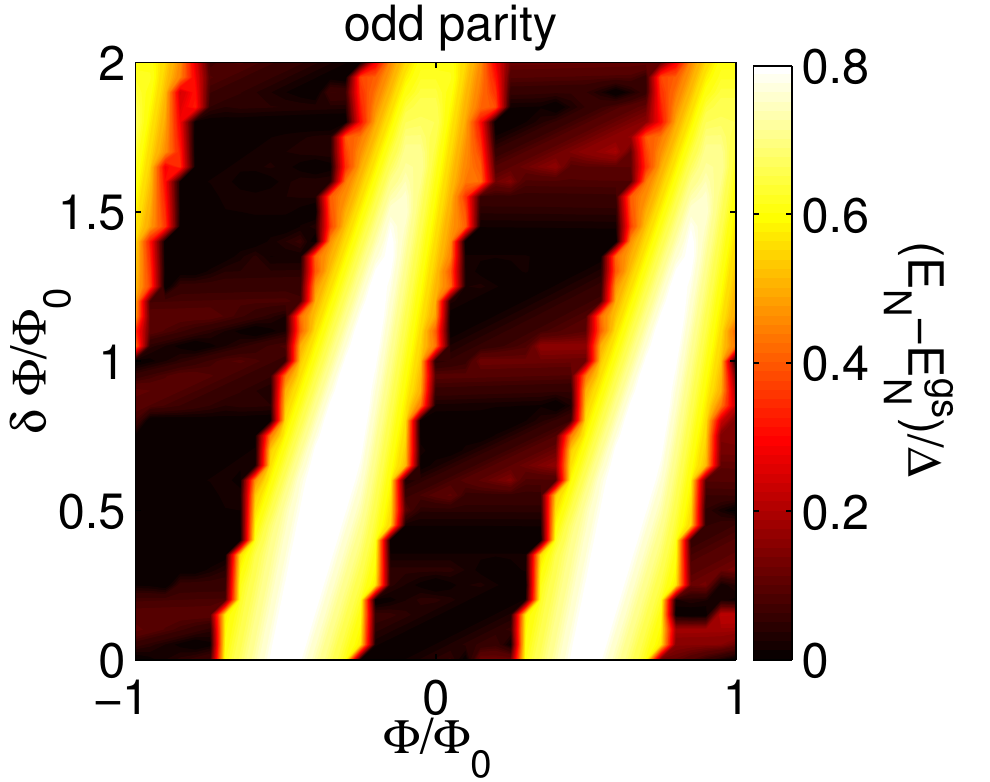}
\end{center}
\caption{Lowest excitation energy for fixed mean particle number as function of magnetic flux $\Phi \, (\mathrm{mod} \, \Phi_0)$ which pierces the nanowire relative to its central line, and as function of additional magnetic flux $\delta \Phi=B L L_\perp/2$ through the nanowire itself due to the finite thickness of the nanowire. The parameters used in the calculation are: odd parity, $N=43$, $E_Z=2\Delta$, $L=2\pi R=3$ $\mu$m, $L_\perp=70$ nm, and $a=10$ nm. For even parity, we find the same spectrum with $\Phi$ shifted by $\Phi_0/2$. } 
\label{fig:flux_shift_multiband}
\end{figure}

Due to the finite width $L_\perp$ of the nanowire, the area of the nanowire itself is penetrated by magnetic flux and thus the magnetic flux through the ring-shaped nanowire is not well-defined. However, the magnetic flux can be decomposed into a mean value for the middle of the wire and deviations due to the finite thickness 
\begin{equation}
\delta \Phi(y)=B L y \,  \, \mathrm{ ~ for} \, -\frac{L_\perp}{2}<y<\frac{L_\perp}{2}.
\end{equation}
For nanowires with radius $R=L/2\pi \sim 0.5\, \mu$m and magnetic field strengths $B \lesssim 1$ T, we find $\delta \Phi(L_\perp/2)>\Phi_0$. In figure \ref{fig:flux_shift_multiband}, we display the fixed electron number excitation spectrum as function of mean magnetic flux $\Phi$ (mod $\Phi_0$) and additional flux $\delta \Phi(L_\perp/2)$. We find that the flux periodicity of the excitation spectrum is not changed, however, the excitation spectrum itself is shifted due to $\delta \Phi$. We can therefore conclude that additional magnetic flux due to the finite width of ring-shaped nanowires with large radii is unproblematic for the study of the flux periodicity of the excitation energies.

%%%%%%%%%%%%%%%%%%%%%%%%%%%%%%%%%%%%
%%%%%%%%%%%%%%%%%%%%%%%%%%%%%%%%%%%%

\section{Sequential and Cotunneling of Cooper pairs}
\label{sec:cotunneling_of_cooper_pairs}

In general, there are competing transport channels through the hybrid ring-shaped nanowire. Our focus is on nonlinear Coulomb blockade transport due to sequential single-electron tunneling through the SM. The most important competing channels are sequential and cotunneling of Cooper pairs. For that purpose, we estimate the magnitude of the current due to Cooper pair processes for a general superconducting island weakly tunnel-coupled to two leads and study under which parameter conditions they become important. We assume that the SC has a large number of transverse channels $N_{\perp}$ and charging energy $E_c$. The tunneling between lead $i$ and the SC is described by
%
%***********************  Coupling Hamiltonian lead island  *****************
\begin{equation}
H_{T,i}=\sum_{\mathbf{k}\mathbf{q}\sigma}\Big\{ t_{\mathbf{k},\mathbf{q}}a^\dagger_{i\mathbf{k}\sigma} \big(u_{\mathbf{q}\sigma}\gamma_{\mathbf{q}\sigma}+ v_{\mathbf{q}\sigma}\gamma^\dagger_{-\mathbf{q}-\sigma}\big)+h.c.\Big\},
\label{eqn:coupling_lead_dot}
\end{equation}
%*************************************************
%
where $t_{\mathbf{k},\mathbf{q}}$ are the tunnel matrix elements, $a_{i\mathbf{k}\sigma}$ are the fermion operators in lead $i$ with energy $\epsilon_{i\mathbf{k}}$, and $\gamma_{\mathbf{q}\sigma}$ are QP operators for the SC with energy $E_{\mathbf{q}}=\sqrt{\xi_{\mathbf{q}}^2+\Delta_{SC}^2}$. The $u_{\mathbf{q}\sigma}$ and $v_{\mathbf{q}\sigma}$ are the BCS coherence factors with magnitudes $\sqrt{(1 \pm \xi_{\mathbf{q}}/E_{\mathbf{q}})/2}$. 

Tunneling of Cooper pairs between lead $i$ and the SC is described by an effective Hamiltonian which can be derived in second order perturbation theory in $H_{T,i}$. In the first step, one electron with momentum $\mathbf{k}_1$ and spin $\sigma$ is transferred from an initial state into an intermediate excited state with momentum $\mathbf{q}$ of the superconducting island. In the second step, another electron with momentum $\mathbf{k}_2$ and spin $-\sigma$ tunnels into the partner state of the first electron $-\mathbf{q}$ such that both electrons form a Cooper pair. Hence, the final state contains an extra Cooper pair in the SC and two QP excitations in the lead. Similarly, we find the reverse process by splitting a Cooper pair followed by two consecutive electron tunneling events \cite{HG1993}. This yields the effective tunneling Hamiltonian
%
%***********************  2nd order perturbation theory  *****************
\begin{equation}
H_{CP,i}= \langle \mathrm{BCS} | H_{T,i} \frac{1}{i\eta -H_0} H_{T,i} |\mathrm{BCS} \rangle ,
\label{eqn:second_order}
\end{equation}
%*************************************************
%
where we traced out the QP operators via the BCS ground state $|\mathrm{BCS} \rangle$. We find
%
%***********************  Hamiltonian Cooper pair tunneling  *****************
\begin{equation}
H_{CP,i}=\sum_{\mathbf{k}_1\mathbf{k}_2}\Big\{ 
A_{i\mathbf{k}_1,\mathbf{k}_2}a_{i\mathbf{k}_1\uparrow}a_{i\mathbf{k}_2\downarrow} + 
A_{i\mathbf{k}_1,\mathbf{k}_2}^*a_{i\mathbf{k}_2\downarrow}^\dagger a_{i\mathbf{k}_1\uparrow}^\dagger \Big\}
\label{eqn:HCP}
\end{equation}
%*************************************************
%
with the effective tunneling matrix elements
%
%***********************  Matrixelements Cooper pair tunneling  *****************
\begin{eqnarray}
A_{i\mathbf{k}_1,\mathbf{k}_2}=\sum_{\mathbf{q}} t_{\mathbf{k}_1, \mathbf{q}}^* t_{\mathbf{k}_2, -\mathbf{q}}u_{\mathbf{q}\downarrow} v_{-\mathbf{q}\uparrow} &  \Big\{ \frac{1}{E_c+E_{\mathbf{q}}-\epsilon_{i\mathbf{k}_1}-\mu_i}  \nonumber \\
&+\frac{1}{E_c+E_{\mathbf{q}}-\epsilon_{i\mathbf{k}_2}-\mu_i} \Big\}. 
\label{eqn:martix_cooper}
\end{eqnarray}
%*************************************************
%

In the following, we consider the Andreev current through a normal-superconducting-normal structure with symmetric barriers and bias voltage $0<V<\Delta_{SC}/e$. Assuming that the voltage between the left (right) lead and the SC is $\pm V/2$, we calculate the rate for the Andreev reflection process using Fermi's golden rule. The current for the scattering of two electrons from the left metallic lead into the SC reads
%
%***********************  current Andreev  *****************
\begin{equation}
I_A(\omega) = 2e\frac{2\pi}{\hbar} \sum_{\mathbf{k}_1 \mathbf{k}_2} |A_{L\mathbf{k}_1, \mathbf{k}_2}|^2  f(\epsilon_{L\mathbf{k}_1})f(\epsilon_{L\mathbf{k}_2}) \delta(\epsilon_{L\mathbf{k}_1}+\epsilon_{L\mathbf{k}_2}+\omega)
\label{eqn:current_Andreev}
\end{equation}
%*************************************************
%
with $\omega=eV-4E_c$ and Fermi functions $f$. In Ref. \cite{HG1993} it has been shown that the Andreev conductance $G_A=I_A/V$ for sequential Cooper pair tunneling can be written as $G_A(\omega)=(e^2/h) G^2(\omega)/N_\perp$, where $G$ is the dimensionless normal state conductance and $N_\perp$ the number of transverse channels through the superconducting region in its normal state. Due to the charge $2e$ of Cooper pairs, sequential tunneling of Cooper pairs is not resonant for $eV/2 < E_c-\Delta_{\mathrm{eff}}$ and can be neglected. In the expression for the current, this suppression shows up as a shifted chemical potential $\omega=eV-4E_c$. 

Similarly, we calculate the current for Cooper pair cotunneling from the left lead to the right lead via the superconducting island by calculating the scattering rate in second order perturbation theory in $H_{CP,i}$. We find 
%
%***********************  current Andreev cotunneling  *****************
%\begin{widetext}
\begin{eqnarray}
I_{A,cot}(V)= 2e \frac{8\pi}{\hbar} \sum_{\mathbf{k}_1 \mathbf{k}_2 \mathbf{k}_3 \mathbf{k}_4} & \frac{ |A_{L\mathbf{k}_1,\mathbf{k}_2}|^2  |A_{R\mathbf{k}_3,\mathbf{k}_4}|^2 }{(\epsilon_{L \mathbf{k}_1}+\epsilon_{L \mathbf{k}_2}+eV-4E_c)^2} \nonumber \\
&f(\epsilon_{L\mathbf{k}_1}) f(\epsilon_{L\mathbf{k}_2})  f(-\epsilon_{R\mathbf{k}_3}) f(-\epsilon_{R\mathbf{k}_4})\nonumber \\
&\delta(\epsilon_{L\mathbf{k}_1}+\epsilon_{L\mathbf{k}_2}-\epsilon_{R\mathbf{k}_3}-\epsilon_{R\mathbf{k}_4}+2eV).
\label{eqn:current_Andreev_cot}
\end{eqnarray}
%\end{widetext}
%*************************************************
%
Building on the result for the sequential Cooper pair tunneling and assuming $eV \lesssim E_c$, we find that the Andreev cotunneling current can be expressed as 
%
%***********************  current Andreev cotunneling 1  *****************
\begin{equation}
I_{A,cot}(V)\approx h  \frac{G_A^2(eV) V^3}{E_c^2}.
\label{eqn:current_Andreev_cot_1}
\end{equation}
%*************************************************
%
In the expression for Eq. \eref{eqn:current_Andreev_cot_1}, the Andreev conductance $G_A(eV)$ is not suppressed by the Coulomb energy since the charge on the superconducting island after the tunneling events is the same as the initial charge. 

In contrast, we find for sequential electron tunneling a current $I_{seq}= (e/h) \Gamma$ where $\Gamma$ is the tunneling rate between the lead and the SM. For characteristic bias voltages smaller or equal to $E_c/e$, we compare the currents due to the sequential tunneling of electrons and the Andreev cotunneling of Cooper pairs. With Eq. \eref{eqn:current_Andreev_cot_1} and the expression for the Andreev conductance, we find 
%
%***********************  current ratio  *****************
\begin{equation}
\frac{I_{seq}}{I_{A,cot}} \approx \frac{N_\perp^2 \Gamma}{E_c G^4}.
\label{eqn:current_ratio}
\end{equation}
%*************************************************
%
We now make the conservative assumption $\Gamma \approx d/10$ and $d \approx E_c/10$, where $d$ is the mean level spacing in the SM, and demand that single-particle sequential tunneling be larger than Cooper pair cotunneling. In this way, we obtain the condition that $G < \sqrt{N}_\perp/3$, i.e. the dimensionless conductance of the junction between lead and the SC in its normal state has to be smaller than one third of the square root of the number of transverse channels. For a metal of diameter 10 nm and with Fermi wavelength 0.3 nm, the number of transverse channels is approximately (diameter/wavelength)$^2 = 1000$, and thus the dimensionless normal state conductance needs to satisfy $G < 10$, which is realistic for metallic quantum dots with current state technology. 

One way to realize the condition $G < \sqrt{N}_\perp/3$ experimentally is to not cover the nanowire with superconducting material in the vicinity of the electrodes. This significantly reduces the conductance between the SC and the electrodes. One can even imagine that an extreme limit could be realized, in which all electrons entering the hybrid system have to do so via the SM in the vicinity of the electrodes. One might argue that as a consequence of removing the SC near the electrodes, the proximity induced pairing amplitude in this region will be reduced as well. However, when the region not covered with superconducting material is considerably smaller than the coherence length in the SM (of the order of 100 nm as shown in Sec. \ref{sec:non_superconducting_segments}), this effect will be small. In principle, one could go even further and only deposit superconducting nanograins on top of the nanowire instead of adding a fully connected SC, and in this way eliminate the influence of Andreev cotunneling almost completely. 

In order to fully suppress cotunneling of Cooper pairs through the SC, we propose to use ferromagnetic leads with the polarization in magnetic field direction. While ferromagnetic leads fully suppress Andreev processes and thus cotunneling of Cooper pairs in conventional $s$-wave SCs, they do not significantly affect the current due to sequential tunneling of electrons.

%%%%%%%%%%%%%%%%%%%%%%%%%%%%%%%%%%%%
%%%%%%%%%%%%%%%%%%%%%%%%%%%%%%%%%%%%

\section{Summary}
\label{sec:summary}

In conclusion, we have proposed a Coulomb blockade transport experiment to investigate the topological order of semiconductor-superconductor hybrid nanorings, and have shown that characteristic parity and flux periodicity effects in the excitation spectrum reflect the distinct ground-state degeneracies of trivial and nontrivial superconducting phases on manifolds with nonzero genus. In particular, the excitation spectrum for fixed mean particle number provides clear signatures of the $h/e$ flux periodicity in the nontrivial phase and the topological phase transition. All these findings are robust against geometry details of the realization of the ring structure and rely on the existence of a hole such that the system is homotopically equivalent to a circle. 

We have shown that the spectroscopic gap in the nontrivial phase is robust against moderate electrostatic disorder. Furthermore, the nontrivial phase is characterized by a large superconducting coherence length which allows to deposit superconducting nanograins on top of the nanowire instead of adding a fully connected superconductor, and in this way reduces the Andreev cotunneling and enhance the charging energy. Using a $T$-matrix formalism, we have estimated the magnitude of Andreev cotunneling and have derived a criterion for the maximum number of parallel conduction channels through the proximity coupled $s$-wave superconductor which ensures that single-particle transport dominates over cotunneling of Cooper pairs. 

Finally, we studied multi-subband nanowires and we have shown that nonlinear Coulomb blockade transport can be used as a tool to map out the topological phase diagram.

\ack

We acknowledge helpful discussion with L. Kimme, and financial support by BMBF.

%%%%%%%%%%%%%%%%%%%%%%%%%%%%%%%%%%%%
%%%%%%%%%%%%%%%%%%%%%%%%%%%%%%%%%%%%

\appendix

\section{Multi-band phase diagram}
\label{sec:multiband_phase_diagram}

\begin{figure}[tb]
\begin{center}
\includegraphics[width=0.8\textwidth ,clip]{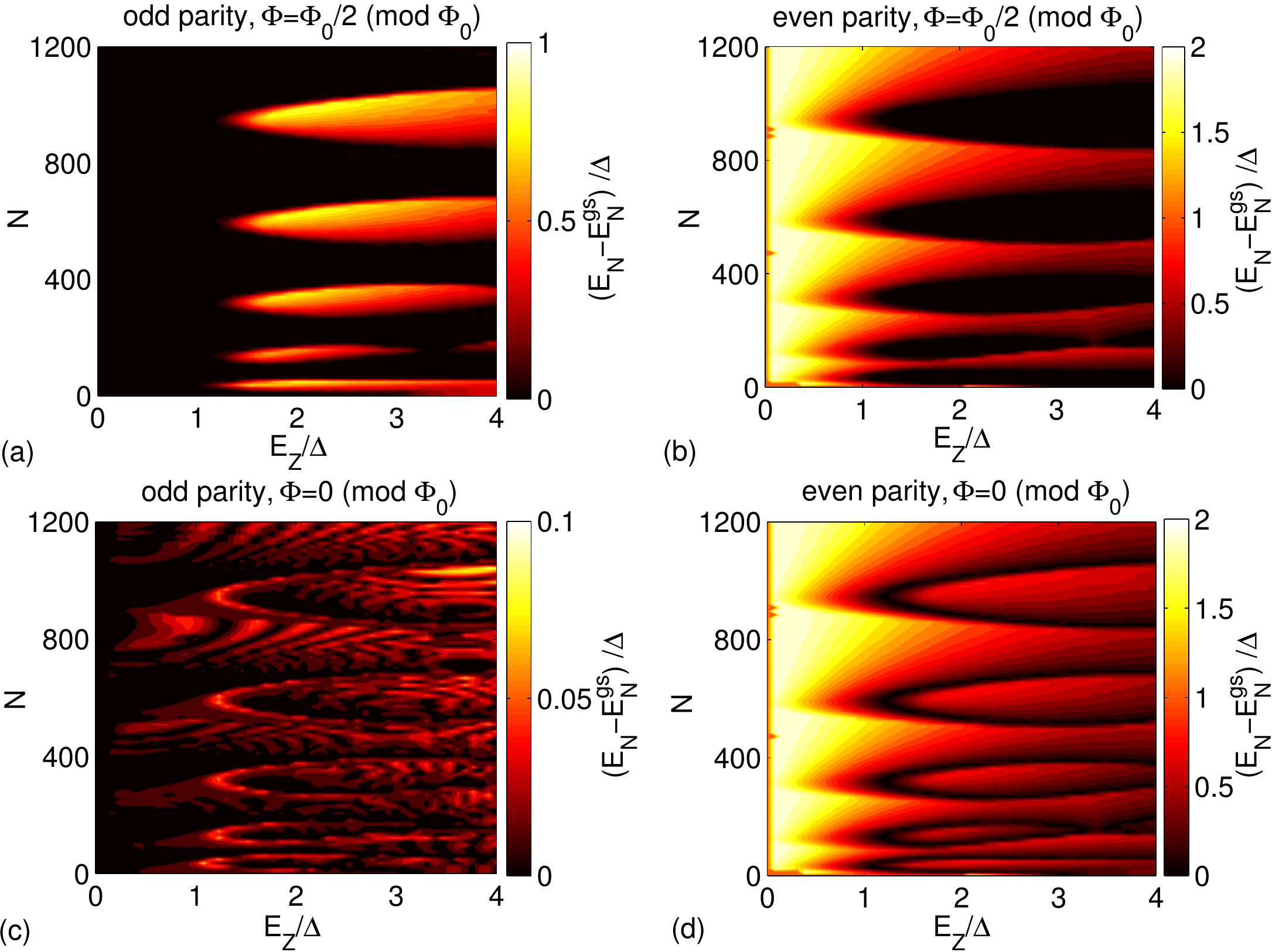}
\end{center}
\caption{Lowest excitation energy in the fermionic excitation spectrum with fixed mean electron number as function of magnetic field and mean electron number for several combinations of magnetic flux and electron parity; $L=3$ $\mu$m, $L_\perp=100$ nm, and $a=5$ nm. Note the different color scale in (c) where all excitation energies are determined by the single-particle level spacing $d$.} 
\label{fig:addition_multiband}
\end{figure}

In this appendix, we present the lowest excitation energy $E_N-E_N^{gs}$ for the multi-band SM hybrid nanowire for several combinations of magnetic flux and parity as function of Zeeman energy and mean electron number. As shown in figure \ref{fig:addition_multiband} and explained in section \ref{sec:multi_band_hamiltonian}, both the chemical potential $\mu$ and the Zeeman energy can be used to tune the nanowire through the topological phase transitions.

Our findings for the trivial phase (i.e. the dark region in figure \ref{fig:addition_multiband}(a)) are characteristic for $s$-wave superconductivity in superconducting grains without excitation gap for odd parity [figures \ref{fig:addition_multiband}(a) and (c)] and with energy gap $2 \Delta_{\mathrm{eff}}$ for even parity [figures \ref{fig:addition_multiband}(b) and (d)]. These excitation energies do not change qualitatively when changing the magnetic flux and show small $\Phi_0/2$ periodic oscillations of order $d^2/\Delta_{\mathrm{eff}} \ll \Delta_\mathrm{eff}$. 

In the topologically nontrivial phase (i.e. the bright region in figure \ref{fig:addition_multiband}(a)) the parity effect is very different. Here, the excitation energies depend on both electron parity and magnetic flux. In Figs.~\ref{fig:addition_multiband}(a) and (d) we find an excitation gap $2\Delta_{\mathrm{eff}}$ since two Bogoliubov QP excitations are required and thus a Cooper pair needs to be broken. In contrast, the excitation energies in Figs.~\ref{fig:addition_multiband}(b) and~\ref{fig:addition_multiband}(c) are determined by the single-particle level spacing since always one unpaired particle is located near the Fermi surface. As shown in Fig.~\ref{fig:addition_multiband}(d), the characteristic signature of the topological phase transition is the closing and reopening of the excitation gap. When studying the flux dependence of the excitation energies in the nontrivial phase, we find large oscillations with period $\Phi_0$ and amplitude $2 \Delta_{\mathrm{eff}}$. For even parity, the excitation energies for $\Phi/\Phi_0 \in (-1/4,1/4)$ are determined by the effective gap $2\Delta_{\mathrm{eff}}$ while they are determined by the single-particle level spacing $d^2/\Delta_{\mathrm{eff}}$ for $\Phi/\Phi_0 \in (1/4,3/4)$. For odd parity, we qualitatively find the same spectrum but shifted by $\Phi_0/2$, as follows from the earlier discussion. 

Thus, the excitation spectrum for fixed electron number directly reflect the topological phase diagram shown in figure \ref{fig:spectrum_multiband}(b). The proposed nonlinear Coulomb blockade transport experiment can therefore be used as a tool to clearly determine the topological order of the hybrid system by measuring the fermionic excitation spectrum.

%%%%%%%%%%%%%%%%%%%%%%%%%%%%%%%%%%%%
%%%%%%%%%%%%%%%%%%%%%%%%%%%%%%%%%%%%

\end{document}